# Hybrid CMOS detectors for the Lynx x-ray surveyor high definition x-ray imager

**Samuel V. Hull,*  Abraham D. Falcone,*  Evan Bray, Mitchell Wages, Maria McQuaide, and David N. Burrows**
Pennsylvania State University, Department of Astronomy and Astrophysics, University Park, Pennsylvania, United States

**Abstract.** X-ray hybrid CMOS detectors (HCDs) are a promising candidate for future x-ray missions requiring high throughput and fine angular resolution along with large field-of-view, such as the high-definition x-ray imager (HDXI) instrument on the Lynx x-ray surveyor mission concept. These devices offer fast readout capability, low power consumption, and radiation hardness while maintaining high detection efficiency from 0.2 to 10 keV. In addition, x-ray hybrid CMOS sensors may be fabricated with small pixel sizes to accommodate high-resolution optics and have shown great improvements in recent years in noise and spectral resolution performance. In particular, 12.5-$\mu$m pitch prototype devices that include in-pixel correlated double sampling capability and crosstalk eliminating capacitive transimpedance amplifiers, have been fabricated and tested. These detectors have achieved read noise as low as 5.4 e$^-$, and we measure the best energy resolution to be 148 eV (2.5%) at 5.9 keV and 78 eV (14.9%) at 0.53 keV. We will describe the characterization of these prototype small-pixel x-ray HCDs, and we will discuss their applicability to the HDXI instrument on Lynx. © *The Authors. Published by SPIE under a Creative Commons Attribution 4.0 Unported License. Distribution or reproduction of this work in whole or in part requires full attribution of the original publication, including its DOI.* [DOI: 10.1117/1.JATIS.5.2.021018]

## 1 Introduction

The Lynx x-ray surveyor[1,2] is one of the four large mission concepts currently under study for the NASA 2020 Decadal Survey, promising great advances for x-ray astrophysics. The baseline mission concept achieves these advances by combining excellent angular resolution (better than 0.5″ on-axis with minimal off-axis degradation) with a large effective area (2.3 m$^2$), resulting in dramatically increased x-ray sensitivity relative to current and planned x-ray missions. This will allow for an extension of x-ray astronomy into the high redshift and low luminosity universe while also providing increased counting statistics and spectral measurements for more nearby sources and for resolved sources for which spatially resolved spectra are critical to scientific advancement. Highlights of specific science goals include studying the birth and growth of the first generation of supermassive black holes, looking for faint absorption lines in the warm-hot intergalactic medium, and studying the impact of feedback on stellar birth and evolution.

One of the three primary science instruments planned for Lynx is the high-definition x-ray imager (HDXI),[3,4] which would provide fine angular resolution imaging and moderate spectral resolution over a wide field-of-view (22′ × 22′). This instrument is baselined to use a large silicon pixel array with sensitivity from 0.2 to 10 keV. In order to oversample the point spread function of the 0.5″ mirrors, HDXI requires pixel sizes $\leq 0.33″ = 16~\mu$m. Initial concept studies also call for HDXI to obtain near Fano-limited energy resolution ($\leq$150 eV at 5.9 keV; $\sim$70 eV at 0.3 keV) and <4 e$^-$ read noise (RMS). Further, high instrument throughput leads to photon pile-up (two or more photons landing in a single pixel before a frame is read out) becoming a major concern and drives a requirement for high-readout speeds, with the current baseline having a requirement of greater than 100 frames/s full frame. This limits the applicability of traditional x-ray CCDs, which would experience saturation for even moderately bright sources at their typically slower readout speeds.[5] Fast readout of x-ray CCDs is possible through the use of higher than normal line speeds and development of devices with many parallel output lines, but increased line speeds will cause significant additional noise and even more power consumption.

X-ray hybrid CMOS detectors (HCDs) are an active pixel silicon sensor technology with many attractive features for HDXI. Featuring in-pixel amplifiers with row–column addressing and the ability to read out individual pixels independently of one another through many parallel output lines, HCDs can read out many times faster than traditional CCDs. Further, the absence of parallel clock gates leads to lower power requirements ($\sim$100 to 200 mW to read out a 1024 × 1024 HCD with associated electronics),[6] and the elimination of charge transfer across many centimeters of silicon grants HCDs increased radiation hardness.[7] Both features are important for future high-throughput x-ray missions, such as Lynx; low power consumption enables the use of large arrays of small pixels operating at high frame rates, and increased radiation hardness will maintain high detector performance over long mission lifetimes. Importantly, HCDs also allow for deep depletion depths and back-illumination, enabling high detection efficiency from 0.2 to 10 keV.

In this paper, we present x-ray HCDs as a potential candidate for the HDXI instrument on the Lynx x-ray surveyor, and in particular, we show results from promising small-pixel HCDs. We begin with a discussion of x-ray HCDs, including the first generation HxRG detectors and the small-pixel HCDs (Sec. 2). We then move to discuss testing of the small-pixel HCDs (Secs. 3 and 4) and present results of measured charge spreading (Sec. 5.1), read noise (Sec. 5.2), gain variation (Sec. 5.3), and energy resolution (Sec. 5.4). Finally, we also discuss simulations on charge cloud size (Sec. 6) and subpixel spatial resolution (Sec. 7).

---

*Address all correspondence to Samuel V. Hull, E-mail: s.hull@psu.edu; Abraham D. Falcone, E-mail: adf15@psu.edu

## 2 X-Ray Hybrid CMOS Detectors

Developed as a joint collaboration between Pennsylvania State University and Teledyne Imaging Sensors (TIS), x-ray HCDs are back-illuminated silicon detectors with a unique two-layer architecture: a silicon absorbing layer and a readout integrated circuit (ROIC) layer. The absorbing layer is responsible for photon-to-charge conversion via photoelectric absorption in the silicon pixel array, and the ROIC acts as a charge-to-voltage signal converter and contains all readout circuitry. The layers are precisely aligned and then "hybridized" by indium-bump bonding each pixel together.

This multilayer approach grants HCDs the ability for separate process optimization. The absorbing layer can use thick, high resistivity silicon for high quantum efficiency (QE) across the soft x-ray bandpass, and the ROIC is optimized for fast readout and on-chip signal processing. Additionally, x-ray HCDs now commonly include a thin layer of aluminum deposited directly on top of the absorbing layer to block optical light since this technique was successfully demonstrated on the initial x-ray HCDs.[8] Figure 1 shows a cross-sectional schematic of an x-ray HCD.

### 2.1 HxRG X-Ray Detectors

The first generation x-ray HCDs were Teledyne HyViSI HAWAII type CMOS detectors known as HxRGs. Here, the "x" in HxRG refers to the size of the ROIC array in multiples of 1024 pixels (1024, 2048, or 4096). These devices are based on the optical/infrared HxRG detectors that have already reached a high level of maturity for astronomy applications.[9,10] These longer wavelength HxRGs are in use in many ground-based telescopes around the world, and in space missions, such as WISE,[11] the Orbiting Carbon Observatory 2 (OCO-2),[12] and the upcoming James Webb Space Telescope.[13] X-ray HxRG HCDs use identical ROICs and bump bonding to these high technology readiness level (TRL) sensors and therefore benefit from the significant effort put into their development. The absorbing layer of the infrared sensors is composed of HgCdTe and is therefore distinct from the silicon absorbing layer of x-ray HxRGs, though silicon HxRGs do have flight heritage, such as on OCO-2. Most recently, an x-ray HxRG has gained flight heritage on a suborbital rocket payload, as discussed below.

The initially developed batch of x-ray HCDs consisted of H1RG detectors with 18-$\mu$m silicon pixel pitch. While these devices demonstrated low dark current ($\sim$0.02 e$^-$/s/pixel at 150 K), moderate read noise (as low as 7 e$^-$),[14] and high QE,[15,16] they suffered from poor spectral resolution due to interpixel capacitance crosstalk (IPC).[17,18] IPC is an unwanted spreading of signal to neighboring pixels due to parasitic capacitance, which arises from the use of source follower amplifiers in the ROIC. In the source follower readout, the charge collected on the input gate of the amplifier is capacitively coupled to neighboring pixels, thereby leading to IPC. These H1RG HCDs were found to have large IPC, with upper limits of 4.0% to 5.5% in the up/down direction, and 8.7% to 9.7% in the left/right direction.

We have also characterized a specially modified engineering grade H2RG detector. This device has a 2048 × 2048 ROIC with 18-$\mu$m pitch bonded to a 1024 × 1024 silicon absorber with 36-$\mu$m pitch. Every absorber pixel has only one ROIC pixel bonded to it, meaning the effective pitch of the hybridized detector is 36 $\mu$m. This nonstandard layout reduces the effects of IPC by increasing the distance between bonded ROIC pixels while using the heritage source follower amplifier ROIC design. The modified H2RG was shown to be a significant improvement and limited IPC to $\sim$1.8%.[19]

We have tested the H2RG with a cryogenic SIDECAR™ application-specific integrated circuit (ASIC). The SIDECAR ASIC, developed by TIS, is used to provide clock and bias signals to HxRG detectors while also performing chip programming, signal amplification, analog-to-digital conversion, and data buffering. Cooling the cryogenic SIDECAR to 200 K results in improved noise performance. Using the cryogenic SIDECAR, this H2RG has been measured to achieve $\sim$6.5 e$^-$ read noise (RMS) and 93-eV resolution [full-width-half-maximum (FWHM)] at 0.53 keV (see Secs. 5.2 and 5.4 for descriptions of read noise and energy resolution).[20] Figure 2 shows an H2RG spectrum with several low energy x-ray lines.

This H2RG detector recently flew on the Water Recovery X-ray Rocket as part of a diffuse soft x-ray spectrometer.[22,23] The detector operated nominally and detected x-rays during flight, raising this particular sensor to TRL-9. Although this x-ray HCD is flight proven, further developments are needed for HDXI due to its fine angular resolution and rapid low-noise readout requirements. More recent generation x-ray HCDs have implemented smaller pixels and increased pixel functionality while also eliminating IPC and improving performance.

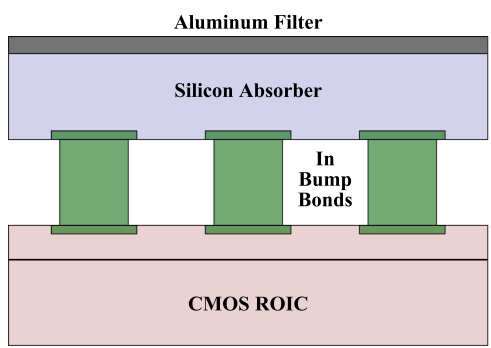

**Fig. 1** Cross-sectional schematic of an x-ray HCD. Si absorbing layer is indium-bump bonded to the readout for each pixel. An aluminum filter is commonly deposited directly on the absorbing layer for optical blocking. Reproduced from Ref. 14.

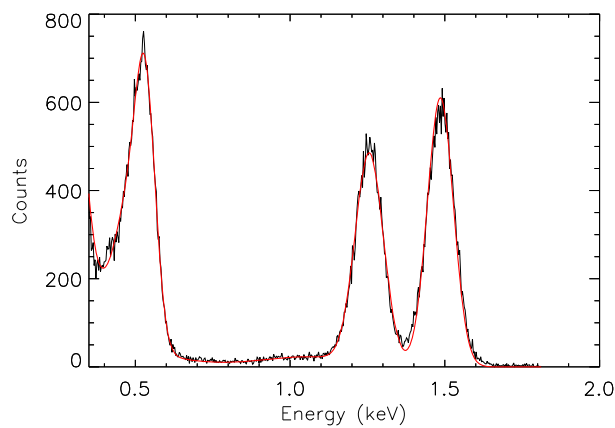

**Fig. 2** An H2RG spectrum with O (0.53 keV), Mg (1.25 keV), and Al (1.49 keV) K$\alpha$ lines. A fit to the data is shown in red. The energy resolution is 93 eV at 0.53 keV, 111 eV at 1.25 keV, and 102 eV at 1.49 keV. Reproduced from Ref. 21.

## 2.2 Small-Pixel HCDs

The small-pixel HCDs are a prototype x-ray HCD designed specifically to meet the needs of future high-throughput x-ray missions with fine angular resolution.[24] The main goal of the small-pixel HCD design was to scale down x-ray HCD technology to smaller pixel sizes while improving read noise and spectral resolution. Our small-pixel HCD test devices are $128 \times 128$ pixel prototypes with 12.5-$\mu$m pixel pitch, and a depletion depth of 100-$\mu$m. Figure 3 shows a small-pixel HCD inside a testing dewar. These test devices, which are initially fabricated in the smaller format to save on development costs, are scalable to multimegapixel size detectors with multiside abuttable packages (typically three-side abutting, as is the case for HxRGs and in-development $1k \times 1k$ small-pixel HCDs).

Each ROIC pixel of the small-pixel HCDs uses a capacitive transimpedance amplifier (CTIA), replacing the source follower of the HxRG detectors. In contrast to the source follower, the CTIA holds the input gate voltage constant during integration and therefore eliminates the problem of IPC. The CTIA was previously implemented on another prototype HCD, the Speedster-EXD, where it was shown to result in unmeasurable IPC (<1%).[25] The small-pixel HCDs also feature the ability to perform in-pixel correlated double sampling (CDS) by subtraction of the variable baseline voltage associated with a pixel reset. This subtraction is carried out on-chip, prior to further amplification.

The small-pixel HCD test devices were designed and fabricated in four "banded arrays" of $128 \times 32$ pixels, by which four distinct pixel designs can be evaluated. The top two bands of each device, or the first 64 rows, contain "type A" pixels that do not include the in-pixel CDS circuitry. The bottom two bands, or the lower 64 rows, contain "type B" pixels with the in-pixel CDS. The upper band of each pixel type contains pixels that include additional guard ring type shielding in the ROIC for guarding against electrical crosstalk. The lower 32 rows of each pixel type contain standard pixels without this extra shielding.

## 3 Experimental Setup

The small-pixel detectors were tested in a modified IR Labs HDL-5 test chamber to expose them to x-rays. The HDL-5 includes a light-tight sealed vacuum section and an unsealed LN2 dewar, which is thermally linked to a cold finger in the vacuum section. The detector is mounted in a socket on a small board in the vacuum section (seen on left in Fig. 3). This board mostly serves to route all external signals while also providing filtering for analog signals. In addition, seven static digital signals (including a signal that determines the pixel type to read out) are set manually by means of jumpers on the board. Two dewar connectors route all lines to an external detector interface board (DIB; developed by TIS) that generates the required power lines and dynamic signals. The detector is configured and read out using the DIB, a Matrox frame grabber, and custom software. The HDL-5 test setup is shown on the right in Fig. 3.

The vacuum section of the HDL-5 is evacuated to pressures of $10^{-6}$ torr, facilitating safe cooling of the detector to near-cryogenic temperatures via filling the LN2 dewar. The detector temperature is controlled using a silicon diode and heater in concert with a Lake Shore temperature controller and was maintained at 150 K. At a typical operating voltage of 15 V applied to the substrate, the detector is fully depleted. Tests were performed at voltages ranging from 15 to 150 V in order to achieve stronger fields and therefore more compact charge clouds within the small pixels.

A radioactive $^{55}$Fe source was used to generate Mn K$\alpha$ and Mn K$\beta$ x-ray lines at 5.9 and 6.49 keV. Two 500 $\mu$Ci $^{210}$Po alpha particle sources were also used to fluoresce a Mg target and produce Mg K$\alpha$ x-rays at 1.25 keV. The Mg target had significant oxidation and was mounted to an aluminum plate that was also exposed to the alpha particles, therefore also generating O K$\alpha$ and Al K$\alpha$ lines at 0.53 and 1.49 keV, respectively.

Four small-pixel HCDs were tested using this setup. These devices, referred to by their serial numbers, are FPA18566, FPA18567, FPA18568, and FPA18569. All four detectors are identical except for the presence or absence of an aluminum optical blocking filter (not expected to impact performance): FPA18566 and FPA18569 have bare silicon absorbing layers with no filter, whereas FPA18567 and FPA18568 have 500 Angstrom aluminum filters deposited directly on their absorbing layers.

## 4 Data Reduction

The HDL-5 test stand was used to obtain charge spread, read noise, pixel-to-pixel gain variation, and energy resolution measurements at a 10-Hz frame rate. Data for type A and type B pixels must be acquired separately due to the different output method for each pixel type. Lacking in-pixel CDS, type A pixels must output both a reset and a signal frame for each exposure, which is time multiplexed within one row time. Our specific DIB firmware implementation captures these type A data as two $128 \times 128$ images that have alternating rows of valid pixel data and empty nondata (zeros). Each reset or signal frame has only 64 rows of valid pixel data (making up the full 64 rows of type A pixels), which is written to even rows for reset frames and to odd rows for signal frames. Each frame was then contracted down to $128 \times 64$ by eliminating zeroed rows, followed by

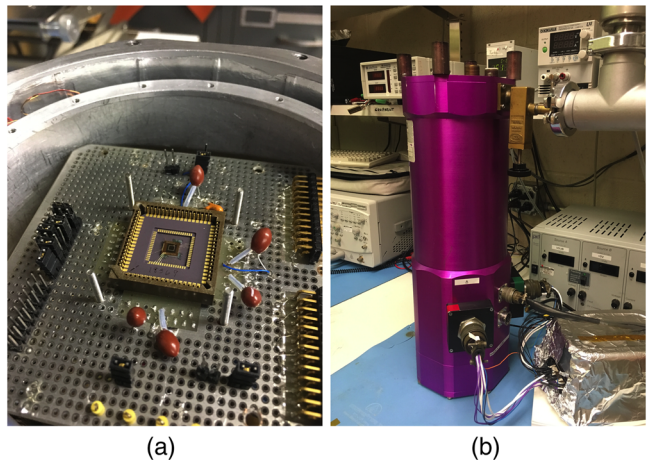

**Fig. 3** (a) Small-pixel HCD inside a testing dewar. (b) The IR Labs HDL-5 dewar test setup. The dewar is 50.8 cm tall and has a diameter of 17.65 cm.

subtraction of the reset frame from the signal frame to create pseudo-CDS images. (Note: these detectors integrate signal down from a high reset level so we then multiply by −1 for the positive signal.)

Type B pixels already perform on-chip CDS and therefore solely output a $128 \times 64$ signal frames for each exposure. Before each data run, a bias frame is constructed from the average of 1000 dark frames without an x-ray source. The bias frame is subtracted from each signal frame to create bias-subtracted type B frames. Bias subtraction is unnecessary for type A frames because we have already directly subtracted a reset frame.

The image subtraction described above removes fixed-pattern noise; however, images may still include horizontal artifacts or "row noise." To remove this nonrandom noise component, a boxcar smoothing algorithm is used that subtracts from each pixel a 21 pixel wide, outlier resistant, moving median value. This results in images that contain only a random pixel noise background.

After boxcar smoothing, we perform x-ray event detection and grading. Event candidates are required to have their primary pixel value above a $5\sigma$ threshold, where $\sigma$ is the read noise (see Sec. 5.2) and be local maxima in their $3 \times 3$ pixel region. A secondary threshold, typically between $1\sigma$ and $3\sigma$, is then used to select any pixel in the $3 \times 3$ region above this threshold to be included in the event and have its signal added to the total event signal. Each event is graded according to the number of pixels included and its shape. Due to the small pixel size and large depletion depth of these detectors, natural charge spreading was found to be quite high—especially at lower substrate voltages—and the acceptable grade list was expanded to include additional event geometries. We use a grading scheme similar to the Swift XRT photon counting mode grading scheme[26] for one and two-pixel events while expanding on allowed x-ray event grades due to the large natural charge spreading in the small pixels. In this paper, we use grade 0 events (all charge contained in a single pixel), grade 1 to 4 events (charge spreads to one adjacent pixel) as well as all event grades consistent with real x-rays. These real event grades can include as many as seven adjacent pixels for low applied substrate voltages but typically include between one and four pixels at higher applied substrate voltages (see Sec. 5.1).

We also choose to exclude any events that fall on a bad or flickering pixel. These are pixels that are either insensitive to x-rays, have large deviations from the mean background level, or spend a significant amount of time at large deviations. The cause of such bad pixels is typically lattice defects that create charge traps with high leakage current or edge effects near the detector boundary. Since these small-pixel HCDs are engineering grade test devices with no screening, they are expected to have a fair percentage of bad and flickering pixels; the exact amount depends on the substrate voltage, as increased edge effects at higher substrate voltages lead to more such pixels. Between 2% and 12% of pixels were excluded during data analysis, except for type A pixels on FPA18567, for which up to 66% of pixels were excluded due to a large region of bad pixels on one corner of this device.

## 5 Analysis and Results

### 5.1 Charge Spreading

As mentioned in Sec. 4, the small pixel size and large depletion depth of these detectors lead to a large amount of natural charge spreading. The charge cloud of an x-ray interacting in the silicon absorbing layer has both a large vertical distance to spread over and a small horizontal distance before reaching nearby pixels, leading to a high percentage of multipixel events. This is typically nonideal because it requires summing charge from multiple pixels to reconstruct an x-ray photon's energy and therefore degrades energy resolution due to including noise from all pixels in the event.

The value of the substrate voltage (VSUB) can have a significant effect on the average number of pixels in an x-ray event. Higher VSUB will provide a stronger electric field that will cause the charge cloud to reach the ROIC sense node with less lateral diffusion into surrounding pixels. If one were aiming to have the best energy resolution and photon response at the softest x-ray energies, one would ideally raise VSUB to a level where the average final charge cloud size is completely contained in the size of a single pixel. However, it should be noted that better subpixel spatial resolution can be achieved through the use of multiple pixel events, so one might use this effect to trade some energy resolution for improved angular resolution using two to three pixel events (see Sec. 7).

FPA18567 was operated at a range of VSUB values from 15 to 150 V while collecting Mn K$\alpha$ and Mn K$\beta$ photons. X-ray events were graded to determine the number of pixels in each event. Events that spread to four to seven pixels dominated images at VSUB = 15 V. However, at VSUB = 100 V, events with five or more pixels were virtually eliminated and there was a dramatic increase in one and two pixel events (see Fig. 4), which dominate the distribution when combined. Results showing the percentage of events with one to four pixels as a function of VSUB are shown in Fig. 5. It becomes practical to select only

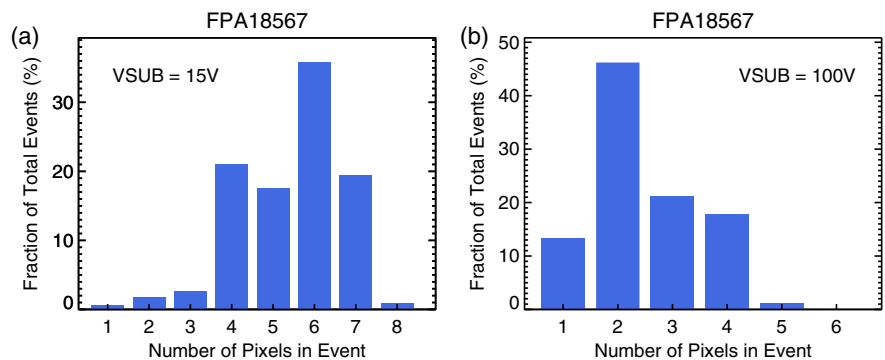

**Fig. 4** Percent of total x-ray events on FPA18567 versus number of pixels included in the event at (a) VSUB = 15 V and (b) VSUB = 100 V.

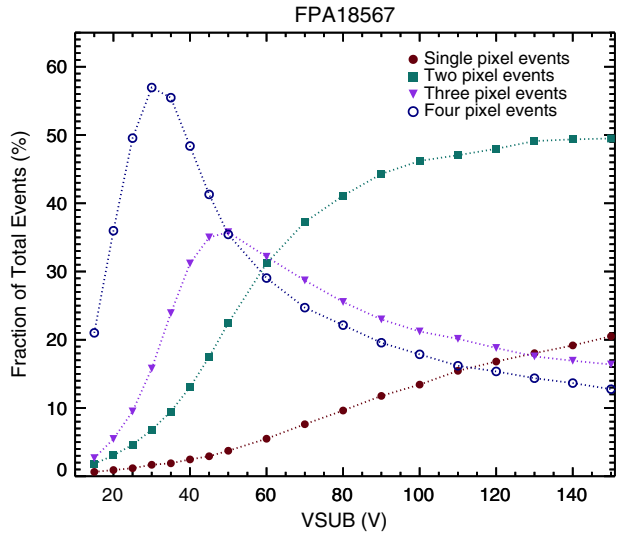

**Fig. 5** The percentage of total x-ray events that include 1 through 4 pixels as a function of VSUB for FPA18567 ($3\sigma$ secondary threshold).

**Table 1** Read noise for each pixel type on small-pixel FPAs. Type A pixels do not include in-pixel CDS while type B pixels do include in-pixel CDS.

| FPA | Shielding | Fit RMS ($e^-$) | Pixel RMS ($e^-$) |
|---|---|---|---|
| **FPA18566** | | | |
| Type B | Extra | $6.5 \pm 0.1$ | $7.4 \pm 2.1$ |
| | Less | $7.3 \pm 0.2$ | $8.6 \pm 2.9$ |
| **FPA18567** | | | |
| Type A | Extra | $5.7 \pm 0.2$ | $7.3 \pm 2.0$ |
| | Less | $5.9 \pm 0.1$ | $7.6 \pm 2.1$ |
| Type B | Extra | $5.8 \pm 0.1$ | $6.8 \pm 2.1$ |
| | Less | $6.9 \pm 0.1$ | $8.2 \pm 2.5$ |
| **FPA18568** | | | |
| Type A | Extra | $5.6 \pm 0.1$ | $7.2 \pm 1.9$ |
| | Less | $5.6 \pm 0.1$ | $7.3 \pm 1.8$ |
| Type B | Extra | $5.4 \pm 0.1$ | $6.3 \pm 2.0$ |
| | Less | $6.1 \pm 0.1$ | $7.1 \pm 2.3$ |
| **FPA18569** | | | |
| Type A | Extra | $5.6 \pm 0.1$ | $6.7 \pm 2.0$ |
| | Less | $5.6 \pm 0.2$ | $6.8 \pm 2.2$ |
| Type B | Extra | $6.1 \pm 0.1$ | $7.0 \pm 2.1$ |
| | Less | $6.8 \pm 0.1$ | $8.0 \pm 2.3$ |

Notes: Fit RMS refers to the noise measurement made by fitting Gaussians to the dark frame histograms, and pixel RMS refers to the noise measurement made by calculating the pixel-by-pixel RMS across all dark frames.

grade 0 and grade 1 to 4 events at VSUB $\approx 50$ V, whereas grade 0 events alone start to become significant at VSUB $\approx 75$ V.

## 5.2 Read Noise

Read noise is associated with the charge-to-voltage conversion step of x-ray detection. It can arise at various points along the readout chain from when charge enters the ROIC sense node to eventual conversion of voltage to a digital value. It effectively sets the noise floor of the detector and also directly impacts the energy resolution.

The small-pixel HCD read noise was measured using two different methods. In each case, 1000 dark exposures that contained no x-ray events were collected for each pixel type and shielding level and then processed with the normal subtraction and boxcar smoothing algorithms. The first method is to fit a Gaussian to each dark frame histogram to determine its standard deviation, followed by taking the mean of these values. The second method is to calculate the mean of the pixel-by-pixel standard deviation across all 1000 dark frames. While both methods characterize the typical RMS deviation of pixel values, the first method (fit RMS) primarily samples the main pixel distribution and is much less sensitive to bad or very noisy pixels compared to the second method (pixel RMS). The fit RMS measurement is, therefore, a more accurate characterization of the real capability of these detectors because a scientific grade HCD should have very few bad pixels, though the pixel RMS method is a more standard noise measurement.

The result of either method will be a measurement of the read noise in units of digital number. The read noise is then converted to units of electrons using the pair creation energy (3.65 eV/$e^-$)[27] and the measured Mn K$\alpha$ peak centroid value. Read noise was measured for all pixel types and shielding levels on FPA18567, FPA18568, and FPA18569. Type B pixels were measured on FPA18566; however, a wire bond on this package was broken before type A data could be collected. Table 1 summarizes all these measurements. The read noise is fairly similar across detectors and pixel types, though type B pixels with less shielding seem to have the highest noise. Measurements tend to be $\sim 5.5$ to $7.5 \, e^-$, with the best measured value equal to $5.4 \, e^- \pm 0.1 \, e^-$.

## 5.3 Gain Variation

HCDs make use of an individual readout amplifier in every pixel of the array, and although every amplifier in the array is designed to the same specifications, the amplifier gain varies slightly across the detector. Unless calibrated and corrected, differences in gain will lead to x-ray photons of the same energy being assigned slightly different energies in different pixels. The net effect will be a degradation of the energy resolution, potentially even dominating over read noise in determining peak width. However, gain can be measured on a pixel-by-pixel basis and therefore, corrections may be applied to account for the variation.

We used the HDL-5 test setup with an $^{55}$Fe source to measure the gain variation of type B pixels for FPA18567, FPA18568, and FPA18569, as well as type A pixels on FPA18569. Due to time constraints, we have not measured all pixel types on all FPAs. Approximately 2500 to 3000 Mn K$\alpha$ grade 0 events per pixel were collected for each detector. Because of the vast amount of data required, event detection, grading, and histogramming were done in real time as images were captured, allowing images to be discarded without saving. In this way, a Mn x-ray spectrum was built up for each pixel in the array, allowing the Mn K$\alpha$ and noise peaks to be fit with Gaussians. The gain was measured in each pixel by subtracting the centroid of the noise peak from the centroid of the K$\alpha$ peak (and multiplying by −1 since these detectors integrate signal down from a high level), and then dividing by the estimate of 1616 e$^-$ per event (assuming $\omega = 3.65$ eV/e$^-$ with $E = 5.9$ keV).

**Table 2** Measured gain variation for FPA18567, FPA18568, and FPA18569.

| FPA | Pixels | Gain variation |
| --- | --- | --- |
| **FPA18567** | B—all pixels | 1.35% ± 0.12% |
| | B—extra shielding | 1.23% ± 0.11% |
| | B—less shielding | 1.44% ± 0.12% |
| **FPA18568** | B—all pixels | 1.18% ± 0.13% |
| | B—extra shielding | 1.12% ± 0.06% |
| | B—less shielding | 1.18% ± 0.07% |
| **FPA18569** | A—all pixels | 1.97% ± 0.13% |
| | A—extra shielding | 1.96% ± 0.07% |
| | A—less shielding | 1.95% ± 0.08% |
| | B—all pixels | 1.24% ± 0.13% |
| | B—extra shielding | 1.16% ± 0.13% |
| | B—less shielding | 1.17% ± 0.14% |

A gain variation map was created for each detector and then the gain variation was calculated as the standard deviation of the gain variation map divided by the mean gain (excluding bad pixels). The results are summarized in Table 2; all three FPAs show gain variation for type B pixels of ~1.2%, whereas the one type A measurement shows gain variation of almost 2.0%. Figure 6

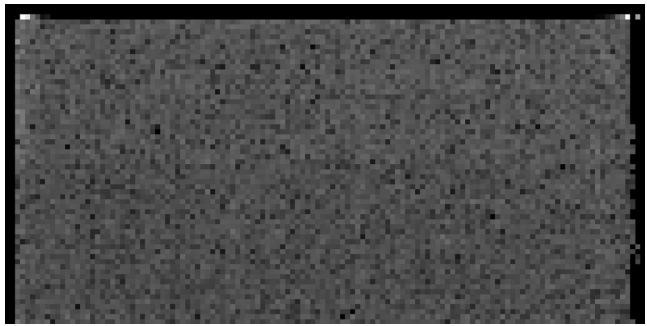

**Fig. 6** Gain variation map for FPA18568 type B pixels (with in-pixel CDS). Brightness of pixels corresponds to the measured gain in that pixel. The measured gain variation is 1.18% ± 0.13%. Black pixels around the sensor edge are insensitive to x-rays due to substrate voltage edge effects and were excluded from the calculation.

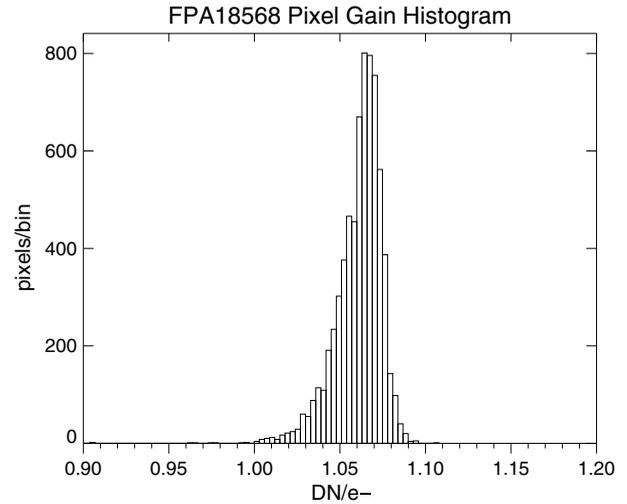

**Fig. 7** Gain histogram for FPA18568 type B (in-pixel CDS) pixels.

shows the calculated gain variation map for FPA 18568, and Fig. 7 shows the pixel gain histogram.

### 5.4 Energy Resolution

Energy resolution ($\Delta E$) is a key detector parameter that determines how well spectral lines can be resolved. For a given spectral line, it is defined as the FWHM of the peak, which is calculated using a Gaussian fit to the line at energy $E$. Energy resolution will often primarily be determined by read noise but also has contributions from dark current, gain variation, and IPC. Assuming negligible dark current and IPC, as is the case for the small-pixel HCDs at 150 K, the energy resolution of a detector (for grade 0 events) is given by

$$\Delta E = 2.354\omega\sqrt{\frac{FE}{\omega} + \sigma^2 + \left(\frac{GV \times E}{\omega}\right)^2}, \qquad (1)$$

where $\omega = 3.65$ eV/e$^-$, $F$ is the Fano factor (0.128 for silicon[27]), $E$ is the incident photon energy, $\sigma$ is the read noise, and $GV$ is the gain variation.

The energy resolution was measured at 5.9 keV for each pixel type and shielding level (again with the exception of type A pixels on FPA18566). Data were taken at VSUB = 15 V using all real x-ray events and with a secondary threshold of $3\sigma$—results are summarized in Table 3. Due to using all x-ray events and therefore summing charge (and noise) from many pixels, energy resolution is somewhat degraded. These results are also presented without any sort of correction for gain variation. We find that type B (in-pixel CDS) pixels with additional shielding were the best performers in all cases, achieving as good as 237 eV (4.0%) energy resolution at 5.9 keV. Although not possible for all devices, with FPA18569, we can correct for gain variation and compare the resolution of types A and B pixels. The data reduction procedure is modified slightly; prior to boxcar smoothing, each image is multiplied by a gain map that converts the image to electron space, after which boxcar smoothing continues and the same steps outlined in Sec. 4 take place. Applying this correction resulted in types A and B pixels achieving comparable resolution for FPA18569, with both pixel types attaining ~240-eV resolution when using all events at VSUB = 15 V.

**Table 3** Energy resolution at 5.9 keV for each pixel type on small-pixel FPAs. Data were taken at VSUB = 15 V and processed with a $3\sigma$ secondary threshold using all real x-ray events. No gain variation correction has been applied.

| FPA | Shielding | $\Delta E$ (FWHM) (eV) |
|---|---|---|
| **FPA18566** | | |
| Type B | Extra | 292 |
| | Less | 398 |
| **FPA18567** | | |
| Type A | Extra | 345 |
| | Less | 343 |
| Type B | Extra | 267 |
| | Less | 421 |
| **FPA18568** | | |
| Type A | Extra | 278 |
| | Less | 288 |
| Type B | Extra | 237 |
| | Less | 325 |
| **FPA18569** | | |
| Type A | Extra | 295 |
| | Less | 311 |
| Type B | Extra | 257 |
| | Less | 317 |

Because they were determined to be the optimal pixel design, all subsequent measurements were carried out using type B pixels with extra shielding. Additional data were taken at higher VSUB on FPA18567 and FPA18568 in order to obtain more one and two-pixel events. Because gain variation maps are available for these detectors, we also apply a gain variation correction. Both detectors collected $^{55}$Fe data with VSUB = 100 V, and energy spectra were created consisting of only grade 0 events, only grade 0 to 4 events, and all real x-ray events. Figure 8 shows a gain corrected spectrum from FPA18568 using only grade 0 events with a $3\sigma$ secondary threshold, which demonstrates 158 eV (2.7%) resolution at 5.9 keV. At this voltage, including all x-ray events only degrades the resolution to 174 eV (2.9%).

If the gain variation had been perfectly accounted for, we could expect to accurately predict the energy resolution using Eq. (1) with $GV = 0$ and the $\sigma$ measured in Sec. 5.2. This results in a theoretical resolution of $\Delta E = 132$ eV (2.2%) at 5.9 keV, which is somewhat better than our corrected result. There is an error associated with measuring the gain map that will factor into this difference, in addition to our measured spectra likely having some events that lost a small amount of charge to neighboring pixels that were not summed into the event, leading to slight peak broadening. Lowering the secondary threshold can help to lessen this effect at the cost of cutting out more events. Lowering the secondary threshold from $3\sigma$ to $2\sigma$, we achieve 148 eV (2.5%) resolution at 5.9 keV, at the cost of going from using 14% of the total events to using 9% of total events. Table 4 summarizes all of the gain variations corrected energy resolution results at 5.9 keV for both FPA18567 and FPA18568.

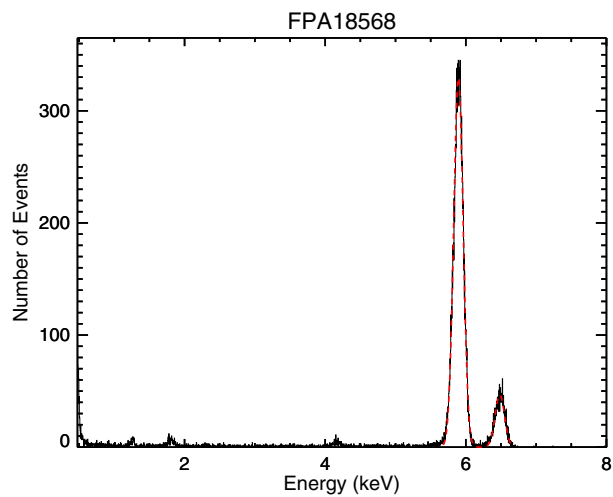

**Fig. 8** Mn K$\alpha$ and K$\beta$ gain-corrected spectrum using only grade 0 events with a $3\sigma$ secondary threshold, from FPA18568 type B pixels with extra shielding. The two Gaussian fit is shown as a dashed red line. The measured energy resolution is 158 eV (2.7%) at 5.9 keV.

A fluorescent magnesium source, producing x-rays at 1.25 keV, was also used to characterize energy resolution. As mentioned in Sec. 3, this setup generated oxygen and aluminum K$\alpha$ x-rays at 0.53 and 1.49 keV, respectively, in addition to the Mg K$\alpha$ x-rays. However, due to low source activity (and quite small collecting area), acquiring sufficient counts required very long times; we, therefore, report results using this source for only one detector—FPA18568. Data were collected at VSUB = 150 V (to maximize single pixel event count rate) and processed with a $2\sigma$ secondary threshold—motivated by previous investigations with HxRGs on the optimal secondary threshold as a function of energy.[19] Figure 9 shows a gain corrected grade 0 event spectrum of all three low energy lines. We achieve FWHM energy resolution measurements of 78 eV (14.9%) at 0.53 keV, 99 eV (7.9%) at 1.25 keV, and 91 eV (6.1%) at 1.49 keV. These are the best measured values to date for an x-ray HCD at these energies. Table 5 shows full results when selecting different event grades; including all events degrades resolution by ∼20 eV. Improving energy resolution for all events is possible by more tightly constraining the charge cloud and/or reducing noise. In the future, we aim to optimize device thickness and substrate voltage (to decrease natural charge spreading) while also working to reduce read noise (see Secs. 6.2 and 8).

**Table 4** Gain variation corrected energy resolution at 5.9 keV for type B (in-pixel CDS) pixels with extra shielding on FPA18567 and FPA18568. Data were taken at VSUB = 100 V.

| | Secondary threshold | | | | | |
|---|---|---|---|---|---|---|
| | $2\sigma$ | | | $3\sigma$ | | |
| FPA | Grades | % of events | $\Delta E$ (FWHM) (eV) | Grades | % of events | $\Delta E$ (FWHM) (eV) |
| **FPA18567** | 0 | 9% | 152 | 0 | 14% | 161 |
| | 0–4 | 48% | 162 | 0–4 | 60% | 169 |
| | All | 100% | 176 | All | 100% | 179 |
| **FPA18568** | 0 | 9% | 148 | 0 | 14% | 158 |
| | 0–4 | 48% | 157 | 0–4 | 60% | 165 |
| | All | 100% | 171 | All | 100% | 174 |

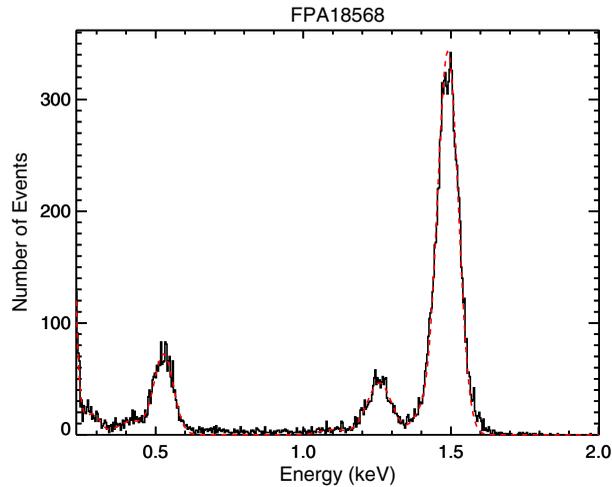

**Fig. 9** Gain-corrected spectrum showing O (0.53 keV), Mg (1.25 keV), and Al (1.49 keV) K$\alpha$ lines from FPA 18568 type B pixels with extra shielding. Spectrum shows only grade 0 events and was processed with a 2$\sigma$ secondary threshold. A fit to the data is shown as a dashed red line. The measured energy resolutions are 78 eV (14.9%) at 0.53 keV, 99 eV (7.9%) at 1.25 keV, and 91 eV (6.1%) at 1.49 keV.

**Table 5** Gain-corrected energy resolution at several low energy x-ray lines for type B (in-pixel CDS) pixels with extra shielding on FPA18568. Data were taken at VSUB = 150 V and processed with a 2$\sigma$ secondary threshold.

| Line energy | Grades | % of events | $\Delta E$ (FWHM) (eV) |
|---|---|---|---|
| 0.53 keV | 0 | 42% | 78 |
| | 0–4 | 84% | 87 |
| | All | 100% | 99 |
| 1.25 keV | 0 | 28% | 99 |
| | 0–4 | 81% | 112 |
| | All | 100% | 117 |
| 1.49 keV | 0 | 25% | 91 |
| | 0–4 | 77% | 103 |
| | All | 100% | 111 |

## 6 Determining Charge Cloud Size

Having experimentally characterized the prototype small-pixel HCDs, we now turn to simulations that explore the implications of changing certain detector parameters. It is first necessary to have a rough understanding of the charge cloud size that is produced by each individual x-ray. Changes in the depletion depth or bias voltage of the detector are the two most important factors in determining charge cloud size, and these will influence the prevalence of different event types as well as the potential for subpixel spatial resolution. Because the current small-pixel prototype devices described in this paper contain too few pixels to be effectively utilized in a mesh experiment (as described in Ref. 28), we explore two other methods for determining the charge cloud size produced by an incident x-ray. The results of these methods are then used to model event type probabilities (Sec. 6.2) and subpixel spatial resolution (Sec. 7) for various absorber layer thicknesses.

### 6.1 Theoretical Treatment

One alternate method for determining charge cloud size is through a series of analytical equations used to describe the process of charge carrier transport in silicon. A full description is outside the scope of this paper but can be found elsewhere.[29–32] These formulae, combined with a model of x-ray interaction depths in silicon, will produce a characteristic distribution of charge cloud sizes for a given set of detector parameters. An example distribution for 5.9 keV x-rays in the current generation of small-pixel detectors is shown in Fig. 10.

### 6.2 Event Type Probabilities

A second method of determining charge cloud size is through direct simulation of the event types produced by an ensemble of x-rays. To do this, a program was written that takes a distribution of charge cloud sizes as an input and uniformly illuminates a pixel with x-ray events. The extent of the charge spreading in the 3 × 3 pixel neighborhood is then determined, and the event is graded with our standard analysis pipeline. By applying a constant scaling factor to the distribution shown in Fig. 10, a charge cloud distribution that best reproduces the observed event type probabilities can be found. It is observed that this method results in a best-fit charge cloud that is ∼25% larger than what is found in Sec. 6.1.

Once a best-fit charge cloud distribution is determined, the detector thickness is varied slightly in order to estimate changes in observed event types. Results are shown in Fig. 11. While a thinner detector will produce more single pixel events due to the reduced drift time of the charge carriers, one must also balance this against reduced QE above 6 keV. Figure 12 shows models for the QE of an x-ray HCD for different thicknesses.

## 7 Modeling Subpixel Resolution

We evaluate several characteristic event types in the following analysis, including single-pixel events, two-pixel split events, and three to four pixel corner split events. These events are predominantly produced by x-rays that land within the center, edges, and corner regions of the pixel, respectively. For this analysis, we assume that all events were produced by a single Mn K$\alpha$ x-ray. A summary of the results for various detector thicknesses is shown in Table 6.

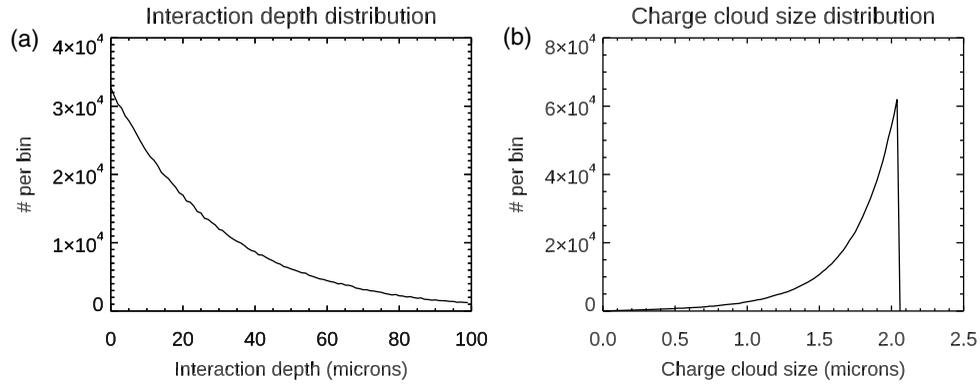

**Fig. 10** An example of the charge cloud size distribution (b) that is calculated from a model of interaction depths (a) in silicon for Mn K$\alpha$ x-rays at 5.9 keV, using an assumed 100 V substrate voltage. X-rays that penetrate further into the absorber layer have less time to diffuse before reaching the collection node. The charge cloud can be well-represented by a 2-D Gaussian, where "size" denotes its standard deviation $\sigma$.

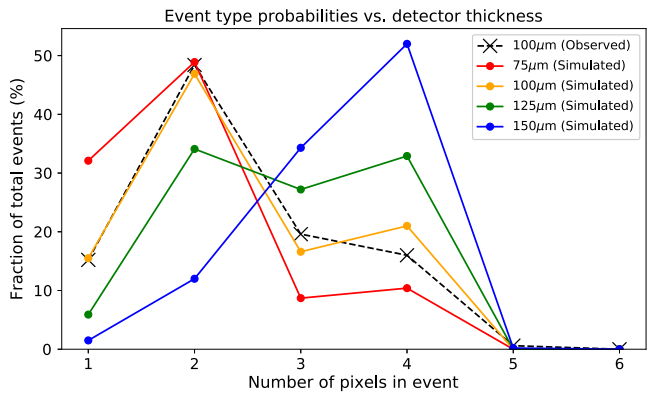

**Fig. 11** An illustration of how event type probabilities for Mn K$\alpha$ x-rays (5.9 keV) can be expected to change as a function of absorber thickness. All cases are for a bias voltage of 100 V. Experimentally determined values are shown as a black "×."

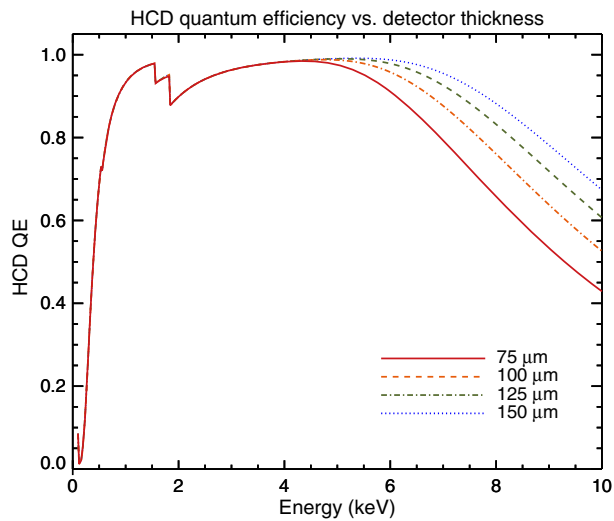

**Fig. 12** Modeled quantum efficiency of x-ray HCDs as a function of energy, shown for different absorber thicknesses. All models include a 500 Angstrom aluminum filter, a 0.25 $\mu$m SiO$_2$ layer, and a 0.1 inactive Si layer.

For single pixel events, some distinction can be made about the landing location of the incident x-ray since it is known that it did not land close enough to a pixel border in order to share a significant amount of charge with neighboring pixels. For these events, this means that the landing location can be restricted to a square region in the center of the pixel with a uniform probability distribution. While this does not offer the far superior subpixel resolution described below for multipixel events, it still represents a much finer spatial resolution than that provided by the individual pixel width alone.

For events that spread a significant amount of charge to neighboring pixels, we utilize a Markov Chain Monte Carlo to fit the charge cloud centroids from the best-fit distribution to a location that best reproduces the observed event. Although knowledge of the centroid location is improved for events that contain a greater amount of signal, it is equally worsened for events where the charge cloud size can take on a broader range of values, as is the case for higher energy x-rays. Illustrations of the best-fit landing location for a horizontal split two-pixel event and for a three-pixel corner split event are shown in Fig. 13. For events that spread charge over two or more pixels, major improvements in spatial resolution can be made by making use of a charge cloud centroid technique. For example, a two-pixel event on a 100-$\mu$m-thick detector with 12.5 $\mu$m wide pixels would typically achieve x-ray centroiding with a 68% confidence region that is $0.2 \times 1.5$ $\mu$m.

## 8 Conclusion

We have presented x-ray HCDs as a potential sensor technology for the HDXI instrument on Lynx while also reporting the characterization of 12.5-$\mu$m pitch small-pixel HCDs specifically designed for this purpose. All pixel types of the small-pixel HCDs have been successfully operated and the in-pixel CDS pixels with extra shielding demonstrated the best performance. We have characterized the amount of charge spreading as a function of applied substrate voltage and find that even with deep depletion depths and small pixels, one can achieve a substantial fraction of one and two pixels events by increasing the substrate voltage to $\gtrsim 50$ V. We also measure read noise to be as low as 5.4 e$^-$ and find the calibratable pixel-to-pixel gain variation to range from 1.1% to 2.0%. The energy resolution was measured, including a correction for gain variation, and was found to be as good as 148 eV (2.5%) at 5.9 keV and 78 eV (14.9%) at

**Hull et al.: Hybrid CMOS detectors for the Lynx x-ray surveyor high definition x-ray imager**

**Table 6** A table summarizing the results of modeled subpixel spatial resolutions at 5.9 keV for various detector thickness and for various landing locations of incident x-rays.

| Detector thickness | 68% Confidence region half-width ($\mu$m) | | | | | | | |
|---|---|---|---|---|---|---|---|---|
| | 75 $\mu$m | | 100 $\mu$m | | 125 $\mu$m | | 150 $\mu$m | |
| X-ray landing location within the pixel | x | y | x | y | x | y | x | y |
| Center | 2.0 | 2.0 | 1.5 | 1.5 | 0.7 | 0.7 | 0.2 | 0.2 |
| Right | 0.2 | 2.0 | 0.2 | 1.5 | 0.2 | 0.6 | 0.2 | 0.2 |
| Bottom | 2.0 | 0.2 | 1.5 | 0.2 | 0.6 | 0.2 | 0.2 | 0.2 |
| Bottom-right | 0.2 | 0.2 | 0.2 | 0.2 | 0.2 | 0.2 | 0.2 | 0.2 |

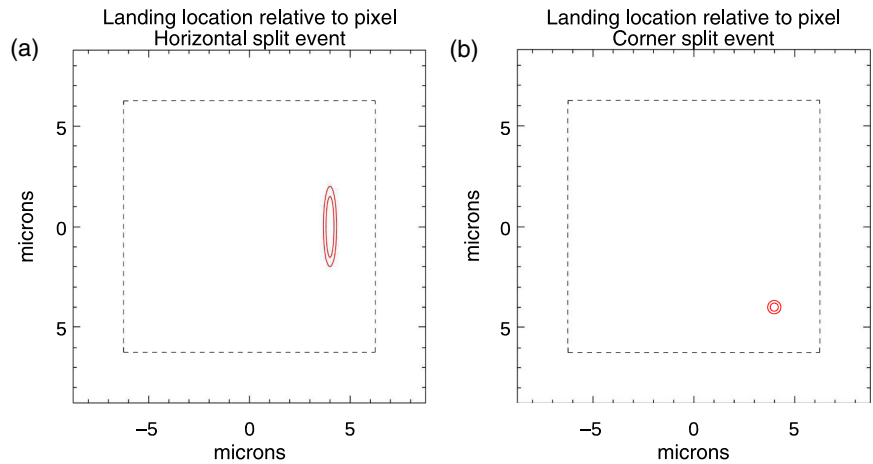

**Fig. 13** The regions within the pixel to which (a) a characteristic horizontal split and (b) three-pixel corner split events can be localized for a small-pixel HCD with a thickness of 100 $\mu$m. The 68% and 90% confidence contours are shown inside of a 12.5-$\mu$m pixel indicated by the dashed line.

0.53 keV (all values are FWHM). Altogether, these measurements make these prototype detectors the best performing x-ray HCDs yet produced, achieving the lowest noise and best energy resolution to date for this technology.

In addition, we briefly explored simulations involving the charge cloud size and subpixel spatial resolution of small-pixel HCDs. With the small pixels being baselined for Lynx, one must optimize the detector thickness to achieve a reasonable number of single pixel events while still maintaining high detection efficiency above 6 keV. At the same time, thicker detectors allow for improved subpixel spatial resolution with better accuracy of charge cloud centroiding techniques. For a 100-$\mu$m-thick detector illuminated by 5.9 keV x-rays, our results show that the 68% confidence region half-widths of the photon landing location can be limited to $\pm 0.2$ $\mu$m in both spatial dimensions for corner split events and to $\pm 0.2$ $\mu$m in one spatial dimension and $\pm 1.5$ $\mu$m in the other spatial dimension for horizontal/vertical split events. Even single pixel events result in greatly improved spatial resolution, limiting the 68% half-widths to $\pm 1.5$ $\mu$m in both dimensions.

The results here are promising for the application of x-ray HCDs for HDXI on the Lynx x-ray surveyor mission. With multiple readout lines, these detectors already meet the fast readout needs of large effective area missions, and the small-pixel HCDs have demonstrated that high performance can be maintained with small pixel sizes. The read noise and energy resolution requirements of HDXI are only slightly more exacting than what these prototypes have demonstrated. Work to further reduce x-ray HCD read noise is an active area of development for future devices. We are also working to develop a large format (at least 1k × 1k) small-pixel detector that includes on-chip analog-to-digital conversion on a row-by-row basis, removing the need for complex supporting readout boards (such as SIDECAR ASICs). This device will use nearly identical pixel ROIC architecture to the test FPAs (with one minor change designed to improve read noise) while implementing a new row–column buffer and overall sensor ROIC design to achieve comparable performance for the full array at 100 frames/s. Other possible future developments could include the addition of event-driven readout, which has been demonstrated on larger pixel HCDs, or other on-chip functionality added to the small-pixel HCDs.

*Acknowledgments*

We gratefully acknowledge support from the NASA APRA Detector Development Program, particularly support from grants NNX13AE57G and NNX17AE35G. We would also like to acknowledge Vincent Douence, Mihail Milkov, Yibin Bai, Mark Farris, Jianmei Pan, James Beletic, and the others at Teledyne Imaging Sensors for their useful discussions and collaboration on past hybrid CMOS developments, as well as for contributions to the operation and design of the small-pixel HCDs.

**Samuel V. Hull** is a graduate student at The Pennsylvania State University, currently working toward a PhD in astronomy and astrophysics. He received his BA degree in physics and mathematics from Washington University in St. Louis in 2015. His research interests include astronomical detectors and high energy astrophysics.

**Abraham D. Falcone** is a research professor of astronomy and astrophysics at The Pennsylvania State University. He leads research in fields ranging from x-ray and gamma-ray instrumentation to high energy astrophysics of active galactic nuclei, x-ray binaries, and gamma-ray bursts. He is a member of the Swift team that was awarded the Bruno Rossi Prize in 2007, as well as a VERITAS collaboration member and a collaborator on multiple future x-ray astronomy mission development efforts.

**Evan Bray** is a graduate student at The Pennsylvania State University, currently working toward a PhD in astronomy and astrophysics. He conducts research on the characterization of x-ray hybrid CMOS detectors and their applications to future x-ray space telescopes. His research includes experimental measurements of subpixel spatial resolution and radiation damage on HCDs.

**David N. Burrows** is a professor of astronomy and astrophysics at The Pennsylvania State University, where he has been involved in x-ray detector development and analysis of data from the ROSAT, ASCA, Chandra, and Swift satellites. He is a coinvestigator on the Chandra Advanced CCD Imaging Spectrometer (ACIS) instrument and is the instrument PI for the Swift x-ray telescope. He was awarded the 2007 Bruno Rossi Prize and the 2009 Muhlmann Award as part of the Swift team.

Biographies of the other authors are not available.